# All-fiber frequency comb at 2 μm providing 1.4-cycle pulses

Sida Xing[1,2]*, Abijith S. Kowligy[1,2], Daniel M. B. Lesko[1,2], Alexander J. Lind[1,2], Scott A. Diddams[1,2]

[1]Time and Frequency Division, NIST, 325 Broadway, Boulder, Colorado 80305, USA
[2]Department of Physics, University of Colorado, 2000 Colorado Ave., Boulder, Colorado 80309, USA
*Corresponding author: sida.xing@colorado.edu

**We report an all-polarization-maintaining fiber optic approach to generating sub-2 cycle pulses at 2 μm and a corresponding octave-spanning optical frequency comb. Our configuration leverages mature Er:fiber laser technology at 1.5 μm to provide a seed pulse for a thulium-doped fiber amplifier that outputs 330 mW average power at 100 MHz repetition rate. Following amplification, nonlinear self-compression in fiber decreases the pulse duration to 9.5 fs, or 1.4 optical cycles. Approximately 32 % of the energy sits within the pulse peak, and the polarization extinction ratio is more than 15 dB. The spectrum of the ultrashort pulse spans from 1 μm to beyond 2.4 μm and enables direct measurement of the carrier-envelope offset frequency using only 12 mW, or ~3.5 % of the total power. Our approach employs only commercially-available fiber components, resulting in a turnkey amplifier design that is compact, and easy to reproduce in the larger community. Moreover, the overall design and self-compression mechanism are scalable in repetition rate and power. As such, this system should be useful as a robust frequency comb source in the near-infrared or as a pump source to generate mid-infrared frequency combs.**

Compact, coherent and broad bandwidth laser frequency comb sources in the mid-infrared (MIR) region (3 μm to 25 μm) are an essential component for molecular spectroscopy, environmental monitoring and other applications [1–3]. While a few frequency comb lasers directly emit in the MIR [4–7], nonlinear frequency conversion from the near-infrared is a general and reliable way to coherently convert mature near-infrared frequency comb sources to MIR wavelengths [8–16]. Among different pump lasers for nonlinear conversion, the 2 μm band with Tm-doped silica fiber holds some unique advantages. The lower photon energy of 2 μm light expands the possibility of nonlinear materials due to reduced multi-photon absorption, and leads to efficient and high power intra-pulse different frequency generation (IP-DFG) [11,17]. And for efficient MIR supercontinuum generation (SCG) in waveguides, the longer pump wavelength also requires less dispersion engineering than telecom band pumps [12]. In addition, frequency comb sources near 2 μm are promising for high precision and sensitive spectroscopy. CO, $CO_2$, $NH_3$ and $CH_4$ have absorption peaks in the 2 μm band [18,19], and direct use of 2 μm light has already attracted interest from the spectroscopy community [20,21]. Finally, the 2 μm wavelength resides in the high transmission window of silica fibers, leading to good availability of commercial components and the same fiber processing techniques as the telecom fibers.

For the generation of ultrashort pulses at 2 μm with thulium-doped fibers (TDF), all-fiber chirped-pulse-amplification (CPA) can deliver multi-watt output [22–24], but it remains challenging to broaden and compress such pulses to the few-cycle region. Compression with free-space gratings has been limited to 80 fs range [22,24], due to the 3rd order dispersion (TOD) and insufficient pulse broadening. To achieve a broadband spectrum, it is essential to perform SCG in a well-controlled manner. In this arena, two main methods are available to yield isolated pulses with octave-spanning spectra: nonlinear self-compression [25], and SCG in normal dispersion fibers [26]. The nonlinear self-compression method combines spectral broadening and compression in one single fiber, making it ideal for implementation in an all-fiber setup. As early as 2007 [25], researchers demonstrated nonlinear self-compression of 2 μm pulses to 17 fs, 0.27 mJ in gas filament. Only very recently [17], 13 fs, 90 nJ pulses were successfully generated using a photonic crystal fiber at 2 μm band.

In this Letter, we present the first sub-2 cycle (9.5 fs) frequency comb source implemented using an all-fiber configuration at 2 μm. By soliton self-frequency shift in highly nonlinear fibers (HNLFs), the seed laser converts the commercially available low noise 1.56 μm frequency comb to 2 μm. A fiber CPA amplifies the 2 μm seed pulse to 420 mW. A two-stage self-compression scheme reduces the pulse duration to 9.5 fs, and outputs 327 mW average power at 100 MHz repetition rate. The all-polarization-maintaining (PM) configuration yields linearly polarized output with 15 dB polarization extinction ratio (PER). Thus, an extra free space polarizer is unnecessary, which is critical for preserving the few-cycle pulse quality. The experimental data show excellent agreement with our simulations in both spectral and temporal domains. From simulations, the 1 μm dispersive wave (DW) has little walk off from the 2 μm pulse allowing us to use only 12 mW of power with an inline f-2f setup to recover 30 dB signal-to-noise ratio (SNR) at 300 kHz resolution bandwidth (RBW). The carrier-envelop offset (CEO) frequency, $f_{ceo}$, has a 3 dB bandwidth of approximately 5 kHz, indicating no degradation of the 1.5 μm seed source. And from the practical side, the whole seed, CPA and compressor can easily fit on a 35 cm x 30 cm breadboard. We only used commercially available components in our implementation, making the laser easily repeatable and accessible.

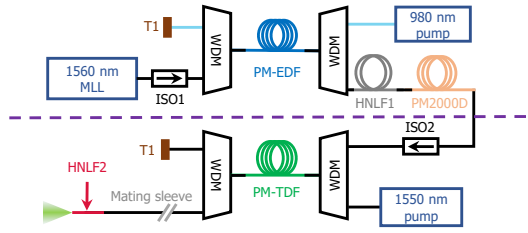

Fig. 1. Experimental setup of the Tm-doped fiber amplifier. Above the dashed line: 2 μm seed and fiber stretcher. Below the dashed line: 2 μm amplifier and compressor. The compression fiber connects with a mating sleeve for better flexibility. MLL: mode-locked laser. ISO: isolator. WDM: wavelength-division multiplexer. T1: terminated port. PM-EDF: polarization-maintaining erbium-doped fiber. HNLF: highly nonlinear fiber. PM-TDF: polarization-maintaining thulium-doped fiber. Black lines are PM1550XP fibers. Cyan lines represent PM980 fibers.

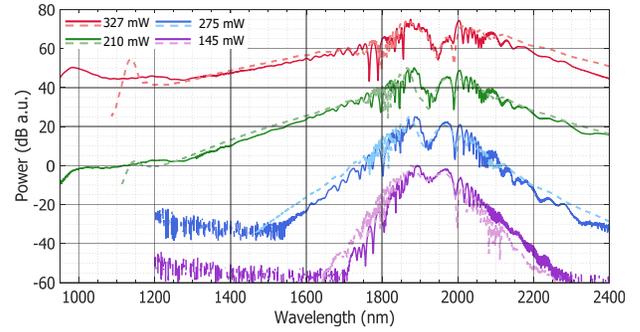

Fig. 2. Simulated spectra (dashed lines) and measured spectra (solid lines) for different powers. For better illustration, each spectrum is shifted by 25 dB. All the simulations share the same fiber parameters.

Figure 1 shows the all-PM fiber configuration for producing 2 μm ultrashort pulses. We begin with a 100 MHz Menlo Systems Figure-9 mode-locked laser that is amplified to 190 mW using a backward pumped nLight PM erbium-doped fiber. The amplified light is then sent into a 30 cm OFS PM HNLF1 to generate a 2 μm pulse by Raman soliton self-frequency shift. The HNLF 1 group velocity dispersion (GVD, $\beta_2$) is -7.82 $ps^2$/km and nonlinear parameter ($\gamma$) is 10 $(W\cdot km)^{-1}$ at 1550 nm. The soliton is centered at 1928 nm with 100 nm bandwidth at 3 dB and an average power of 40 mW. We then use 3 m of Nufern PM2000D to stretch the soliton and ensure net normal dispersion inside the CPA. We estimated the dispersion of PM2000D using [27] to be 92 $ps^2$/km. The stretched pulse then passes a fast-axis blocked PM-isolator (ISO2) with 0.8 dB insertion loss. The isolator blocks backward amplified spontaneous emission (ASE) and filters the light on the fast-axis, and the PER following ISO2 is measured to be greater than 20 dB. A PM-WDM, with <1 dB insertion loss from 1840 nm to 2100 nm, combines the 2 μm seed and continuous-wave 1550 nm pump light. This provides close to 30 mW of 2 μm seed for the PM-TDF, which consists of 2 m of Nufern PM-TSF-9/125 fiber. The forward pumping scheme favors amplification in the 1900 nm region, and the output PM-WDM has > 20 dB isolation between 1.55 and 2 μm to efficiently remove the residual C-band pump. The output of the amplifier connects to the compression fiber through a narrow-key FC/APC mating sleeve for flexibility in optimizing the compression stages, but it could be spliced instead.

For efficient pulse compression, we utilized a 2-stage compression scheme. In the first stage, Nufern PM1550XP fiber compensates the net normal dispersion of the CPA. The optimal fiber length is found to be 64 cm from a cutback test. After the first nonlinear compression stage, the pulse duration is near-transform limited at about 90 fs. The second compression stage utilizes nonlinear self-compression inside an elliptical core PM-HNLF from OFS (HNLF2 in Fig. 1). The HNLF2 is spliced to the PM1550XP using a $CO_2$ laser splicer, with combined splicing and Fresnel reflection losses of 1.2 dB. The strength of the nonlinear compression depends on the power from the thulium amplifier. At the shortest pulse duration, the 2 μm output power is 327 mW with a 1550 nm pump power of 1.94 W, yielding close to 17 % efficiency. Since thulium has more efficient pump absorption at 1.6 μm [28], we expect higher pump efficiency if a 1.6 μm pump is used. After HNLF2, the output is collimated using an off-axis parabolic (OAP) mirror for pulse characterization and other applications. Using a broadband (1000 nm to 2000 nm) polarizer, the compressed pulse PER is measured to be more than 15 dB.

Since the laser outputs an octave-spanning spectrum, two optical spectrum analyzers (OSA) were used to record the complete spectrum. A Yokogawa AQ6375 recorded spectra in the region of 1200 nm to 2400 nm, and when the laser output power was above 270 mW, the spectrum below 1200 nm was recorded using a Yokogawa AQ 6370. The spectral resolution is 2 nm in both OSAs. The solid lines in Fig. 2 show the recorded spectra at four output powers ranging from 145 mW to 327 mW, with step size of 1dB to 1.5 dB. The amplifier output power from 145 to 330 mW corresponds to 1.7W to 1.94 W pump power. Empirically [10,29–31], a structured central region of the spectrum with smooth wings covering about one octave is a general indication of successful self-compression. A more detailed study of pulse quality, duration and temporal distribution requires use of generalized nonlinear Schrodinger equation (GNLSE) with the inclusion of Raman and shock terms. Numerically, implementing the GNLSE from [32], we simulated the complete process from the 2 μm seed soliton to the self-compressed output. The HNLF2 GVD is about -13 $ps^2$/km and nonlinear parameter is 3.1 $(W\cdot km)^{-1}$ at 1930 nm. The fiber parameters used in the simulation are the same for all output power levels (Fig. 2). Due to the short length of HNLF2 (6 cm), the linear propagation loss in silica fiber can be safely ignored even for the 2.5 μm part of the spectrum [25].

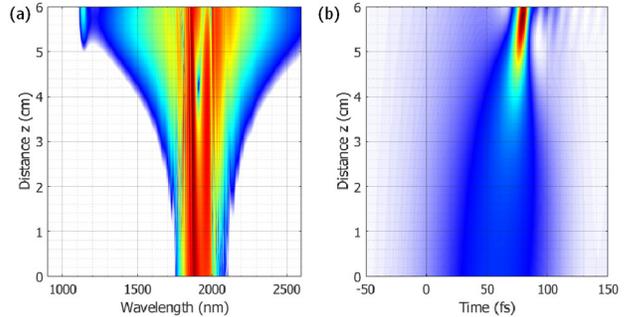

Fig. 3. Simulation of pulse propagation in HNLF2. The output of the first compression stage is used as an input for this simulation (a) Spectral evolution along the fiber length (color in log scale); (b) Corresponding temporal evolution along fiber length (color in linear scale).

The optimal 6 cm length of HNLF2 to produce the shortest pulse is influenced by both the TDF gain center and the self-compression length. The TDF gain center moves to shorter wavelength with increased C-band pump power. For the best SNR, it is beneficial to set the CPA output spectrum peak to overlap with the seed pulse peak. In this case, we found the optimal C-band pump power is around 1.9 W, which gave an output power of 327 mW. Figure 3 shows the simulated pulse evolution along HNLF2 under these conditions, with the simulated output pulse from first stage compression being used as the input pulse for the simulation. The pulse evolution of Fig. 3 indicates that most of the pulse broadening happens within the last 1 cm propagation of HNLF2, with the maximum at about 6 cm. In Fig. 3(b), we show the corresponding temporal evolution of the pulse. Due to the short final pulse duration, we zoom in at the -50 fs to 150 fs region. At the end of HNLF2, the simulated pulse duration is 9.2 fs, with a portion of the laser output power going into satellite pulses. From the simulation, we estimate that the main peak contains 35% of the total power, leading to a peak power of 110 kW.

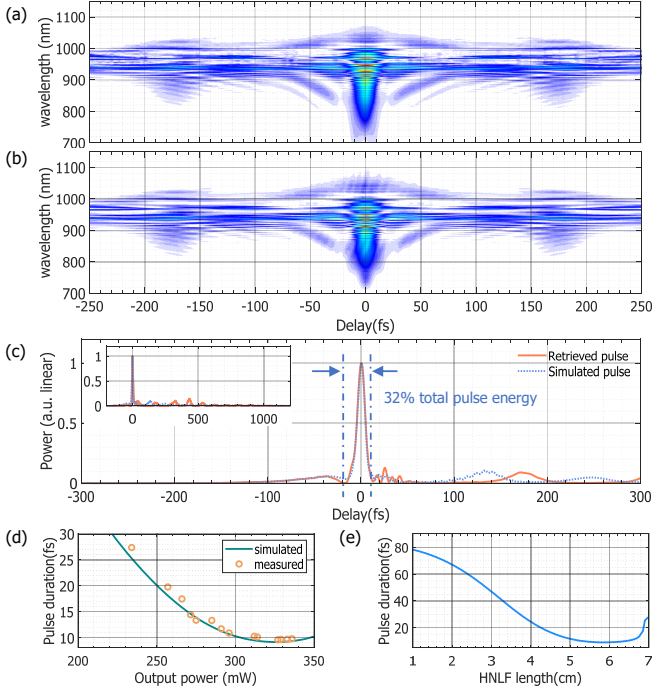

Fig. 4. (a) Experimentally recorded SHG FROG spectrogram. (b) Reconstructed FROG spectrogram, with reconstruction error of 1.1%. (c) Retrieved pulse (solid line) and simulated pulse (dotted line). Inset: retrieved pulse (solid line) and simulated pulse (dotted line) for a delay range is -200 fs to 1200 fs. (d) FROG retrieved pulse duration (dots) and simulated pulse duration (line) as a function of laser output power. (e) Simulated pulse duration as it propagates along the HNLF2.

We perform pulse characterization using a purpose-built second-harmonic generation frequency resolved optical gating (SHG-FROG). In Fig. 4(a) and (b), we show the recorded and reconstructed FROG spectrograms. The FROG scan range is ±600 fs, and the reconstructed spectrogram has an error of 1.1 %. In Fig. 4(c), we zoom at ±300 fs for better comparison between experiment and simulation. Here we see excellent agreement between the simulated pulse duration of 9.2 fs and the reconstructed pulse of 9.5 fs. To check for the presence of satellite pulses, the scan range is increased to ±1.2 ps as shown in the inset of Fig. 4(c). Since there is no pulse component measured beyond -200 fs, we only plot the range from -200 fs to 1200 fs. Even at this range, the simulation still matches well with the experiment. Using the data from ±1.2 ps scan range, the main peak contains about 32 % of total power, hence 104 kW in the peak, with 10 % loss due to Fresnel reflection at the end of HNLF2. Using the estimated GVD and nonlinear parameter, the compressed pulse soliton number (N) is about 1.4. The simulation indicates further increase of C-band pump power (even by a few percent) leads to longer pulses. We measured the pulse duration at various laser powers [dots in Fig. 4(d)]. Just as the simulation matches the measured spectra for various laser powers (Fig. 2), the retrieved pulse duration matches the simulation at all laser powers. In Fig. 4(e), we show the simulated pulse duration as a function of HNLF2 length. To generate around 9.5 fs pulses, we can tolerate a ±2 mm error on fiber length. Such precision is achieved using a normal ruler.

With an octave-spanning spectrum available from the soliton self-compression, it is possible to retrieve the $f_{ceo}$ in a straightforward manner [34]. For high SNR on $f_{ceo}$, the second harmonic (SH) and DW should have good temporal overlap. Checking the pulse evolution simulation in Fig. 3(a), one can notice the 1 µm DW in our laser builds a moderate power (same height as the 1.5 µm edge) at about 5.8 cm. Even though the dispersion difference between the DW and central 2 µm pulse is large, we expect very small walk-off induced by 2 mm of propagation. In addition, the SHG of the 2 µm pulse will fall right onto the DW pulse peak. Therefore, we expect to have an $f_{ceo}$ with high SNR by directly beating the SH of the 2 µm pulse with the DW, using no extra pulse delay control.

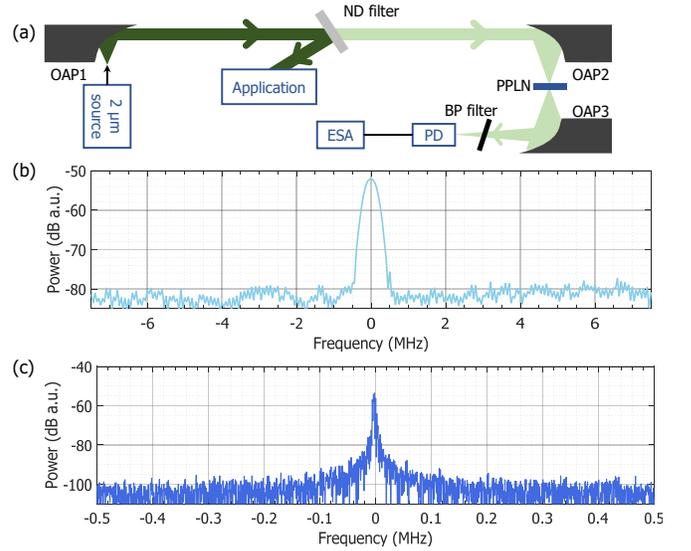

Fig. 5. (a) Experimental setup for $f_{ceo}$ retrieval. A reflective neutral-density (ND) filter is a dispersion-free beam splitter for the reflected beam. The arrows are the beam propagation direction. PPLN: periodically-poled lithium niobate. BP filter: bandpass filter. PD: photodiode. ESA: electrical spectrum analyzer. OAP: off-axis parabolic mirror. (b) $f_{ceo}$ at 300 kHz RBW, a frequency offset of 57.5 MHz is applied. (c) $f_{ceo}$ at 1 kHz RBW, with the same 57.5 MHz offset.

The experimental setup for $f_{ceo}$ measurement is drawn in Fig. 5(a). If the full laser power hits the PPLN without the ND filter, we can easily achieve 34 dB SNR at 300 kHz resolution bandwidth (RBW). However, such demonstration has limited practical applications – it is more interesting to use the minimum power for $f_{ceo}$ detection. Therefore, a partially-reflective metallic neutral density filter (Thorlabs continuous variable NDC-100C-4M filter) is used to separate a 3.5% of the few-cycle pulse, which is focussed into a 1 mm PPLN with poling period of 30 µm. An angle-tuned bandpass (BP) filter at (1075 ± 25) nm passes the DW and SHG light onto the detector (EOT-3000A), and the $f_{ceo}$ heterodyne spectrum is recorded on an electrical spectrum analyzer. At 300 kHz RBW, 13 mW power after the ND filter is enough to achieve >30 dB SNR, as indicated in Fig. 5(b). Figure 5(c) shows the $f_{ceo}$ spectrum at 1 kHz RBW and the 3 dB bandwidth is approximately 5 kHz, which is consistent with the linewidth of the Er:fiber MLL. Such SNR is high enough for a stable frequency comb locking. Meanwhile, about 96 % of the total light can be CEO-stabilized and ready for other applications. It is worth mentioning that the setup in Fig. 5(a) should lead to better $f_{ceo}$ SNR and less power requirement after further optimization. One potential improvement is to use a BP filter centered at 990 nm to utilize higher SHG and DW power (see the DW in Fig. 2).

In summary, we demonstrated and characterized an all-fiber 2 μm frequency comb that outputs 9.5 fs (1.4-cycle) pulses at 100 MHz repetition rate and 15 dB PER. The average output power is 327 mW under 1.94 W C-band pump, and about 32 % of the total power is contained within the main pulse peak. Improving the pulse quality from the first compression stage will contribute to higher power in the central pulse. Thanks to the octave-spanning spectrum and small pulse walk off, direct measurement of $f_{ceo}$ requires only 3.5 % of the total laser power. Most of the laser power can go to other applications, like IP-DFG, SCG or spectroscopy. Our result represents the first sub-2 cycle 2 μm frequency comb in a compact, robust and efficient all-fiber configuration. Importantly, the use of all commercially available components makes our laser configuration repeatable and reliable. This setup is also scalable in repetition rate, which would lead to more average power in each comb line. For example, we anticipate that adapting our approach to all-PM commercially-available heavily-doped Tm-fiber will lead to sub-2 cycle 2 μm lasers working at GHz repetition rates [35]. At the same time, MW level peak powers at a 100 MHz repetition rate should also be feasible with our all-fiber configuration. And finally, even shorter pulses could be feasible through coherent pulse superposition with our 2 μm source, and other regions of the spectrum also generated from the 1.5 μm seed [36].

This work is supported by NIST, the DARPA SCOUT program and AFOSR (FA9550-16-1-0016). The mention of specific companies, products or tradenames does not constitute and endorsement by NIST. The authors thank Esther Baumann and Nima Nader for their comments on this paper and Henry Timmers for his assistance in early stages of this experiment.

**Disclosures.** The authors declare no conflicts of interest.